\documentclass[11pt]{article}
\usepackage[left=3.17cm,top=2.54cm,right=3.17cm,bottom=2.54cm,nohead,nofoot]{geometry}
\usepackage{setspace}
\usepackage{amsmath}
\usepackage{amsfonts}
\usepackage{amssymb}
\usepackage{graphicx}
\newtheorem{thm}{Theorem}[section]
\newtheorem{prop}[thm]{Proposition}

\newtheorem{defi}[thm]{Definition}
\newcommand{\pf}{{\bf Proof. \ }}
\newcommand{\qed}{\hfill $\Box$ \\}
\date{}
\begin{document}
\title{A Practical Quantum Bit Commitment Protocol}
\author{S. Arash Sheikholeslam\\
sasheikh@uvic.ca
\and  T. Aaron Gulliver\\
 agullive@ece.uvic.ca 
\\
Department of Electrical and Computer Engineering, University of Victoria,\\
PO Box 3055, STN CSC, Victoria, BC Canada V8W 3P6}
\maketitle
\newpage
\begin{abstract}
In this paper, we introduce a new quantum bit commitment protocol which is practically secure against entanglement attacks.
A general cheating strategy is discussed and shown to be practically ineffective against the proposed approach.
\end{abstract}
\section{Introduction}
Quantum cryptography in the sense of key distribution was first introduced in \cite{1984-175-179} with the BB84 protocol.
The authors also proposed a bit commitment scheme which they determined was not secure.
Construction of an unconditionally secure quantum bit commitment technique has since become an important research problem.
There have been many commitment schemes created, as well as a number of results on the impossibility of secure commitment \cite{PhysRevLett.78.3414} \cite{Mayers99thetrouble} \cite{Lo:quant-ph9711065}.
Even teleportation has been considered to achieve unconditional security \cite{	arXiv:quant-ph/0305142v3}.
Recently, ``practically'' secure commitment schemes \cite{Quantum Information Processing} have been examined, rather than asymptotically secure protocols.

Consider a two party (Alice and Bob) bit commitment.
Alice chooses a bit $b \in \{0,1\}$, locks it and sends it to Bob (this is called the commitment phase).
When it is time to reveal $b$ (opening phase), Bob locks the bit with his own lock (i.e., he locks the bit locked by Alice),
and sends it back to Alice.
She then opens her lock and sends the bit back to Bob and announces $b$.
Bob then opens his lock and checks whether the locked bit $b$ is the same as the one which was announced.

Here we propose a simple scheme using the principles of the well-known Diffie-Hellman key exchange protocol (details of this protocol can be found in \cite{Menezes:1996:HAC:548089}).
However, we employ multiplication by a unitary transform instead of exponentiation in a multiplicative group modulo a prime.
Although this commitment scheme also falls within the category for which entanglement cheating is a proof of insecurity,
(since it satisfies the criteria based on the simplified Yao model \cite{yaomodel} as described in \cite{Lo:quant-ph9711065}),
it is practically very hard for Alice to cheat.
This is due to the fact that building the unitary transform required to apply on her share of the entangled pair
is practically infeasible, as will be shown.

Before presenting our bit-commitment protocol, we first define practical security.
For this, we need the following.
\subsection*{Binding Experiment (BE)}
\begin{itemize}
\item Alice and Bob share a system $H_A \otimes H_B$ and a protocol $\Pi$ for which the final state before the opening phase is $\rho_{AB} \in H_A \otimes H_B$
\item A cheating Alice performs the operation $A\otimes I[\rho_{AB}]$ and reveals $b\leftarrow_R \{0,1\}$ to Bob. ($A$ is a trace preserving operation)
\item Bob then performs the operation (actually a measurement) $I\otimes B[\rho_{AB}]$ to obtain $b'$.
\item The outcome of the experiment is 1 (success) if $b=b'$ and 0 (fail) otherwise.
\end{itemize}
\begin{defi}
A protocol $\pi$ is computationally binding (CB) if for all polynomial time quantum operations Alice can perform we have
$\Pr[BE_{\pi}^A(1^n)=1] \leq \frac{1}{2} + negl(n)$, where $negl(n)$ is a negligible function of the secrecy parameter $n$.
\end{defi}
\begin{prop}
If a protocol is CB then there is no collection of circuits $\{Q_x \vert x \in S\}$ (where $S$ is any string) which can be generated in polynomial time that can approximate the operation $A$.
\end{prop}
\pf
The proof is obvious given the definition.
\qed
Achieving CB security is a general task and Alice may employ different approaches in an attempt to compromise the security of a protocol.
One important case is an EPR attack by Alice.
EPR attacks \cite{Lo:quant-ph9711065} have been proven to make all quantum bit commitment schemes theoretically insecure.
Therefore we introduce the notion of EPR-Computationally Binding (EPR-CB).

\begin{defi}
A protocol $\pi$ is EPR-Computationally Binding (EPR-CB) if for all polynomial time quantum operations by Alice,
we have $\Pr[BE_{\pi}^A(1^n)=1] \leq \frac{1}{2} + negl(n)$, where $negl(n)$ is a negligible function of the secrecy parameter $n$.
Note that Alice is only capable of entangling an ancillary system in the corresponding Hilbert space,
and can perform unitary transforms and POVM(Positive Operator Valued Measure) measurements on her part before the opening phase.
\end{defi}

\begin{prop}
CB is equivalent to EPR-CB if a cheating Alice can extend any system to a larger system in polynomial time.
\end{prop}
\pf
Obviously, any EPR-CB protocol is also CB.
It is known that all trace preserving quantum operations on a Hilbert space can be extended to a higher dimensional system
in which these operations can be reduced to a unitary transform.
Therefore, a cheating Alice can extend a system and then perform a unitary transform.
A general CB experiment on a Hilbert space $H^n$ is equivalent to a (unitary and POVM)-CB experiment on a Hilbert space $H^m$ where $m \geq n$.
Therefore EPR-CB security is equivalent to CB security.
\qed
Note that this proof is important as it connects the concept of binding to EPR security.

\begin{defi}
An ensemble of protocols $\Pi=\{ \pi_1, \cdots, \pi_n\}$ is computationally binding (CB) if all $\pi_i \in \Pi$ are CB.
\end{defi}
This definition is needed because if there is only one protocol for which the bit commitment is CB,
a cheating Alice can prepare the necessary circuit for changing the qubit in advance and use it at the time of commitment.

\section{The Proposed Bit Commitment Protocol}

In this section, we present the proposed method of bit commitment.
With this protocol, each party prepares a secret unitary operator.
It is assumed that a quantum channel as well as a classical side-channel are available, as with other bit commitment schemes.
The qubits are exchanged through the quantum channel, while the side-channel is used to exchange the secret unitary operators in the opening phase.
The proposal can then be described as follows.
\begin{itemize}
\item Commitment Phase:
\begin{itemize}
 \item Bob prepares two previously agreed upon orthogonal states $\vert\phi_0\rangle, \vert\phi_1\rangle$, and applies his secret transform $U_B$ on them.
  He sends these to Alice and tells her which to use if she wants to commit $0$ or $1$.
\item Alice prepares $U_A \cdot U_B\vert\phi_0\rangle$ or $U_A \cdot U_B\vert\phi_1\rangle$ and sends $\vert\phi\rangle \in \{ U_A \cdot U_B\vert\phi_0\rangle, U_A \cdot U_B\vert\phi_1\rangle \}$ back to Bob depending on the bit she wants to share.
\end{itemize}
\item Opening Phase:
\begin{itemize}
 \item Alice reveals her unitary transform $U_A$ to Bob through the classical channel.
\item Bob computes $\vert \psi \rangle = U_B \cdot U_A\vert\phi\rangle$ and checks if it agrees with the committed qubit.
\end{itemize}
\end{itemize}
Note that the secret unitary transforms can be chosen at random from a continuous subset of the unitary group.
As an example, we can assume that $\vert \phi_0\rangle= \vert 0\rangle$ and $\vert \phi_1\rangle= \vert 1\rangle$,
and $U_A, U_B \in \{R_x(\theta), R_y(\theta), R_z(\theta)\}$ where $R_x(\theta)$ is a rotation about the $x$ axis with an angle $\theta$.

\section{Security and Cheating Strategies}


One approach for Alice to attempt to cheat is to apply a unitary transform $U_A$ during the committing phase but then send $V \cdot U_A$ during the opening phase
(where $V$ is another unitary transform), such that when Bob tries to open the commitment he receives a bit other than the one which was committed
(say Alice has committed $\vert\phi_0\rangle$ but now wants Bob to open $\vert\phi_1\rangle$).
For Alice to be successful in cheating, the following must be true for the last step of the opening phase
\[
\vert \psi\rangle=U_B \cdot V \cdot U_A \cdot U_A.U_B\vert \phi_0 \rangle=\vert\phi_1\rangle \Rightarrow  U_B\cdot V\cdot U_B=\vert\phi_1\rangle\langle\phi_0\vert
\]
This shows that Alice can construct such a transform $V$ only if she knows the secret transform of Bob.
By a similar analysis, Bob also cannot determine the state $\vert \phi_i\rangle$ if he knows
$U_A \cdot U_B\vert \phi_i\rangle$.

\subsection{Practical security against an EPR (entanglement) attack}

Let $\vert A\rangle$ and $\vert B\rangle$ denote the uniform superposition of all possible $U_A$ and $U_B$ on $\vert \phi_i\rangle$.
In other words, assuming $U_A$ and $U_B$ are controlled gates and $\vert A\rangle$ and $\vert B\rangle$ the corresponding control registers, we have a register
($\vert A\rangle$ or $\vert B\rangle$) which is a superposition of all possible choices of the unitary transformations by Alice and Bob.
Considering these registers at the end of the commitment phase, we have
\[
\begin{array}{l}
\vert \psi_0 \rangle=\sum_A \sum_B \vert B\rangle U_A U_B \vert \phi_0 \rangle \otimes U_A U_B \vert \phi_1 \rangle \vert A\rangle\\
\vert \psi_1 \rangle=\sum_A \sum_B \vert B\rangle U_A U_B \vert \phi_1 \rangle \otimes U_A U_B \vert \phi_0 \rangle \vert A\rangle,
\end{array}
\]
where $\vert \psi_0 \rangle$ denotes 0 and $\vert \psi_1 \rangle$ denotes 1.
In each state, the component on the right side of the tensor product is possessed by Alice.
Now, if the protocol is secure against Bob then the local trace over the system components of Alice must be equal for both $\vert \psi_0 \rangle$ and
$\vert \psi_1 \rangle$.
As a result, regarding the Schmidt decomposition we have a unitary transform $V$ on Alice's side which can take values from
$\vert \psi_0 \rangle$ to $\vert \psi_1 \rangle$.
In order for Alice to produce $V$, she must know all possible choices for $U_B$ (but she does not need to know a particular choice of $U_B$).
The existence of $V$ shows that the protocol is not theoretically secure, but the two parties can hide their sets of unitary transforms and make the
protocol practically secure against an entanglement attack.

\section{Conclusions}

In this paper, we proposed a simple but secure bit commitment protocol which is based on the application of
secret unitary transforms by each party (Alice and Bob) in succession.
Cheating strategies, including entanglement cheating, were examined and the system was shown to be effective against these attacks.


\begin{thebibliography}{9}

\bibitem{1984-175-179}
Bennett, Charles H and Brassard, Gilles:
{Quantum cryptography: Public key distribution and coin tossing}, volume 11, Proceedings of IEEE International Conference on Computers Systems and Signal Processing, Bangalore, India, 1984, 175-179

\bibitem{PhysRevLett.78.3414}
{Unconditionally Secure Quantum Bit Commitment is Impossible},
{Apr},
{Phys. Rev. Lett.},
{10.1103/PhysRevLett.78.3414},
{Mayers, Dominic},
  {1997},
  {17},
{http://link.aps.org/doi/10.1103/PhysRevLett.78.3414},
{American Physical Society},
{3414--3417},
{78}

\bibitem{arXiv:quant-ph/0305142v3}
arXiv:quant-ph/0305142v3

\bibitem{Quantum Information Processing}
Ariel Danan, Lev Vaidman: {Quantum Information Processing}, (1 September 2011), pp. 1-7

\bibitem{Menezes:1996:HAC:548089}
 {Menezes, Alfred J. and Vanstone, Scott A. and Oorschot, Paul C. Van},
 {Handbook of Applied Cryptography},
 {1996},
 {0849385237},
 {1st},
 {CRC Press, Inc.},
 {Boca Raton, FL, USA}

\bibitem{Mayers99thetrouble}
{Dominic Mayers}
{The Trouble with Quantum Bit Commitment}
{Computing Research Repository (CoRR)}
{1999}

\bibitem{Lo:quant-ph9711065}
{H. -K. Lo and H. F. Chau}
{Why Quantum Bit Commitment And Ideal Quantum Coin Tossing Are Impossible}
{1997}
{quant-ph/9711065}
{10.1016/S0167-2789(98)00053-0}

\bibitem{yaomodel}
{Yao, Andrew Chi-Chih}
{Security of quantum protocols against coherent measurements}
{Proceedings of 1995 ACM Symposium on Theory of Computing}
{(May, 1995)}
{67-75}
\end{thebibliography}
\end{document}